\documentclass[aps,prb,twocolumn,groupedaddress]{revtex4-1}
\pdfoutput=1

\usepackage{amsmath}
\usepackage{amssymb}
\usepackage{graphicx}
\usepackage{bm}
\usepackage{times}
\usepackage{hyperref}
\hypersetup{%
  breaklinks = {true},
  citecolor = {blue},
  colorlinks = {true},
  linkcolor = {red},
  pdfauthor = {\textcopyright\ Johan Nilsson},
  pdfcreator = {\LaTeX\ and \flqq hyperref\frqq},
  pdffitwindow = {true},
  pdfmenubar = {true},
  pdfpagelayout = {SinglePage},
  pdfstartview = {Fit},
  pdftoolbar = {true},
  plainpages = {false},
}

\newcommand{\Hca}{\mathcal{H}}

\newcommand{\la}{\langle}
\newcommand{\ra}{\rangle}
\newcommand{\ua}{\uparrow}
\newcommand{\da}{\downarrow}

\usepackage{color}
\usepackage{color}

\newcommand{\e}{\epsilon}

\begin{document}


\title{Faithful fermionic representations of the Kondo lattice model}

\author{Johan Nilsson}
\affiliation{Department of Physics, University of Gothenburg, 412 96 Gothenburg,  Sweden}
\email{johan.nilsson@physics.gu.se}

\date{June, 2011}

\begin{abstract}

We study the Kondo lattice model using a class of canonical transformations that allow us to faithfully represent the model entirely in terms of fermions without constraints. The transformations generate interacting theories that we study using mean field theory. Of particular interest is a new manifestly O(3)-symmetric representation in terms of Majorana fermions at half-filling on bipartite lattices. This representation suggests a natural O(3)-symmetric trial state that is investigated and characterized as a gapped spin liquid.

\end{abstract}

\maketitle

\section{Introduction}

One of the standard models in strongly correlated condensed matter physics is the Kondo lattice model, for reviews see e.g. Refs.~\onlinecite{Hewson_Book,Sigrist_RMP_1997,Coleman2002,Gulacsi2004}. The basic physics that this model aims to describe is the simultaneous existence of and interaction between conduction $c$-electrons and localized $f$-spins. It is challenging to describe both subsystems on equal footing because the algebras of the creation and annihilation operators of the $c$-electrons and the $f$-spin operators are quite different. The aim of this work is to introduce and study faithful fermionic representations of this model.

There exists many different approximation schemes that have been used to study this model, for example Gutzwiller projection,\cite{Rice_Ueda_1985,Brandow1986} slave particle methods,\cite{Coleman1984a} large-N approximations,\cite{Read_Newns_1984,Coleman1983} canonical transformations,\cite{Ostlund2007}  mean field theories,\cite{Lacroix_Cyrot_1979,Newns_Read_1987}  and dynamical mean field theory.\cite{DMFT_RMP1996} For discussions of these methods and more references we refer to the reviews just cited. Numerous numerical studies have also been performed. These are particularly successful in one dimension (1D),\cite{Sigrist_RMP_1997} but also the two-dimensional (2D) Kondo lattice have been studied using quantum Monte Carlo at half-filling where the sign problem is absent.\cite{Assaad1999}

All approximate treatments have their shortcomings. The validity of large-N approximations are questionable when one considers the physical case of $N=2$ for example. Slave-particle representations are popular because they keep the spin and charge symmetries clearly visible. One drawback of this approach is that the exact constraint that should be imposed is typically only imposed on average and not exactly. This can be improved at the expense of introducing gauge fields.\cite{Lee_Nagaosa_Wen_RMP} In this work we use exact canonical transformations and hence this drawback will not affect us.

Another physically appealing approximation scheme involves a simple mean field decoupling, see e.g. Ref.~\onlinecite{ZhangYu2000}. This is found to be in qualitative agreement when compared to more involved numerical methods.\cite{Assaad2010} One problem with this approximation is that it does not reproduce the correct size of the Hilbert space of the Kondo lattice. Because the mean field Hamiltonian has two bands for each spin component, the size of the Hilbert space is $2^{4}$ per unit cell. This is appropriate for the periodic Anderson model when the onsite interaction strength $U$ is small. In the Kondo lattice there are $2^3$ states per unit cell however. This problem will not affect us in this work since we will construct faithful representations using three fermions per unit cell.

As we will see there is considerable freedom in the allowed canonical transformations. Of particular interest is a new manifestly O(3)-symmetric representation on bipartite lattices at half-filling. This representation is most easily described in terms of Majorana fermions. One way to motivate it is to start from the known representation of a spin-$\frac{1}{2}$ in terms of three species of 
Majorana fermions: $S_f^i = -i \e^{ijk} \mu_j \mu_k /2$. This representation has a long history that goes back to the fifties, see e.g. the references in Ref.~\onlinecite{Tsvelik_book}. Similar representations have become increasingly popular in recent years after the introduction of the Kitaev model.\cite{Kitaev2006} This representation was also heavily used in earlier works focusing on non-Fermi liquid behavior in modified Kondo impurity problems,\cite{Coleman1995,Bulla1997} as well as lattice systems,\cite{Coleman1995} and odd-frequency pairing in Kondo lattice models.\cite{Miranda1994} The fact that this representation can be used to study Heisenberg models and generate spin liquid states has also been known for some time.\cite{Tsvelik1992,Shastry_Sen_1997,Tsvelik_book} A variant of such a spin liquid state on the triangular lattice was suggested very recently.\cite{Biswas2011}

It is also well appreciated that the representation of a spin-$\frac{1}{2}$ in terms of three Majorana fermions is redundant. In fact the operator $\gamma_0 = 2 i \mu_1 \mu_2 \mu_3$ commutes with the spin operator and can be viewed as an independent Majorana fermion. The O(3) representation can be obtained by writing the $c$-electron creation and annihilation operators in terms of three other independent Majorana fermions and this composite Majorana fermion $\gamma_0$. The Heisenberg exchange interaction between the $f$-spin and the spin of the $c$-electrons then takes on a simple form with manifest O(3)-symmetry. To the best of our knowledge this procedure has not been used previously to describe the standard Kondo lattice model.

This construction can be straightforwardly generalized to the Kondo lattice. But the local O(3) symmetry is a result of the underlying SO(4) symmetry of the Hubbard model at half filling on bipartite lattices,\cite{Yang_Zhang_1990} and its generators are combinations of spin and pseudospin generators (see Sec.~\ref{sec:O3}). It is therefore only possible to generalize the local O(3) symmetry to a globally O(3)-symmetric Hamiltonian on bipartite lattices. On bipartite lattices it is then easy to write down a simple translationally invariant O(3)-symmetric mean field state. This state is found to be favorable when the kinetic term is small or moderately large compared to the exchange interaction. The resulting state is a gapped spin liquid with nonzero triplet pairing amplitude.

The paper is organized as follows: For completeness we end this introduction by writing down the standard form of the Kondo lattice Hamiltonian explicitly. In Sec.~\ref{sec:canonical} we discuss canonical transformations for the 1-site Anderson model and derive different representations for this model. In particular we consider the limit of this model that reproduces the 1-site Kondo impurity problem. These local transformations can easily be generalized to the lattice since they involve canonical fermions. In Sec.~\ref{sec:O3} we provide an elementary derivation of the O(3)-symmetric representation of Sec.~\ref{sec:canonical} using Majorana fermions. We also consider the hopping of $c$-electrons. On bipartite lattices the hopping term can be chosen in a particularly symmetric form with a manifest global O(3)-symmetry. The resulting Hamiltonian is written down explicitly in Eq.~\eqref{eq:Htbsame} of Sec.~\ref{sec:meanfield} where we study this interacting fermion model using variational mean field theory at half-filling. The variational solutions are worked out in 1D for simplicity. Of particular interest is an O(3)-symmetric mean field state that is found to be favorable for some parameter values. This spin liquid state is characterized further in  Sec.~\ref{sec:characterization} where it is shown to have rotationally invariant spin-spin correlations as well as a nonzero triplet pairing amplitude. Our conclusions and an outlook are to be found in Sec.~\ref{sec:conclusions}, and some mathematical details in an Appendix.

\subsection{The Kondo lattice model}

In this paper we study the Kondo lattice model that can be described by the Hamiltonian\cite{Sigrist_RMP_1997}
\begin{multline}
\label{eq:H_KLM}
H_{\text{KLM}} = -t \sum_{\sigma} \sum_{\la i, j \ra} 
c^\dagger_{c,\sigma}(\bm{r}_i) c^{\,}_{c,\sigma}(\bm{r}_j)
+ \text{h.c.}
\\
- \mu \sum_{i} \bigl[n_c(\bm{r}_i) -1 \bigr]
+ J \sum_{i} \bm{S}_c (\bm{r}_i) \cdot \bm{S}_f (\bm{r}_i).
\end{multline}
The first line describes the hopping of $c$-electrons on the lattice, and $\sigma=\, \ua,\da$ is  a spin label. For simplicity we consider only hopping between nearest neighbor lattice sites $\bm{r}_i$ and $\bm{r}_j$, as indicated by the notation $\sum_{\la i,j \ra}$.
The first term on the second line can change the average number of $c$-electrons away from half-filling by introducing a non-zero chemical potential $\mu$.
$n_c(\bm{r}_i) = \sum_\sigma c^\dagger_{c,\sigma} (\bm{r}_i) c^{\,}_{c,\sigma} (\bm{r}_i)$ is the number operator of the $c$-electrons.
The second term on the second line describes the interaction between the spins of the $c$-electrons and the localized $f$-electron spins. On each site of the lattice, $\bm{S}_f (\bm{r}_i)$ is a local spin-$\frac{1}{2}$ object (we use units such that $\hbar=1$ throughout the paper) that satisfies the usual spin-$\frac{1}{2}$ algebra $[S^i , S^j] = i \e^{ijk} S^k$ and $\bm{S}^2 = 3/4$.\cite{Sakurai_QM} The spin operator of the $c$-electrons can be represented as
$\bm{S}_c (\bm{r}_i) = \frac{1}{2} \sum_{\sigma,\sigma'}
c_{c,\sigma}^\dagger (\bm{r}_i) \bm{\tau}_{\sigma,\sigma'} c_{c,\sigma'}^{\, } (\bm{r}_i)$, with $\tau^a$ ($a=1 \ldots 3$) the standard Pauli matrices.\cite{Sakurai_QM}

\section{Canonical transformations for the 1-site Anderson model}
\label{sec:canonical}

The spin-spin interaction term in \eqref{eq:H_KLM} can be derived from the periodic Anderson model in the limit that the occupation of the $f$-electrons is not fluctuating. For this derivation it is sufficient to consider a 1-site Anderson model on each site of the lattice. This is the model that we consider in this section and we extend the formalism in Ref.~\onlinecite{Ostlund2007} to allow for more general canonical transformations. Because we are dealing with only one site we will suppress the site index in this section.

\subsection{1-site Anderson model}
\label{sec:transformation_eh}

We consider first the symmetric 1-site Anderson Hamiltonian at half filling
\begin{equation}
H_{1\text{A}} = -W \sum_{\sigma=\da,\ua} (c^\dagger_{c,\sigma} c^{\,}_{f,\sigma} + c^\dagger_{f,\sigma} c^{\,}_{c,\sigma}) + U (n_f-1)^2 ,
\label{eq:H_Anderson1}
\end{equation}
where $n_f = \sum_\sigma n_{f,\sigma} = \sum_\sigma c^\dagger_{f,\sigma} c^{\,}_{f,\sigma}$. The Hilbert space of this model consists of 16 states that are enumerated in Eq.~\eqref{eq:statesenumeration_0}. Following Ref.~\onlinecite{Ostlund2007} we then introduce another basis using electron $e_\sigma^{\dagger}$ and hole $h_\sigma^{\dagger}$ creation operators, the corresponding states are enumerated in Eq.~\eqref{eq:statesenumeration_s}.

We define the states in the new basis according to the following rules: 1) The singlet ground state $ |0\ra_s$ of Eq.~\eqref{eq:H_Anderson1} is equal to the vacuum state $|1\ra_{eh}$ that has no quasi-particles (i.e. no electrons or holes). 2) The lowest energy eigenstates with the appropriate charge and spin (measured with respect to the singlet ground state) are mapped to the states with one quasi-particle, i.e. $h_\da^\dagger |1\ra_{eh}$, $e_\da^\dagger |1\ra_{eh}$, $h_\ua^\dagger |1\ra_{eh}$, and $e_\ua^\dagger |1\ra_{eh}$. 3) The assignment of the other states is fixed by the charge and spin quantum numbers, except for 4) the two remaining singlets that can be assigned by considering the parity transformation that exchanges the $f$- and $c$-electrons. 
This uniquely defines the quasi-particle operators up to a gauge choice. We fix this arbitrariness by demanding that the quasi-particles go smoothly to the bonding and antibonding eigenstates in the non-interacting limit $U =0$. The transformation matrix that implements this transformation is given explicitly in Eq.~\eqref{eq:Umat_eh}.

This transformation is similar to the one in Ref.~\onlinecite{Ostlund2007}, except that it provides an adiabatic connection between the low-energy states in the non-interacting and the interacting systems. Since the quasi-particle excitations also have the same spin and charge quantum numbers as the non-interacting particles this transformation is an explicit demonstration of the Landau quasi-particle concept.

\subsection{Kondo limit $U \rightarrow \infty$, fermionic representations}

The Kondo limit can be obtained by taking $U,W \rightarrow \infty$ while keeping $J = 4 W^2/U$ finite. In this limit the Hamiltonian \eqref{eq:H_Anderson1} becomes
\begin{equation}
H_{1\text{A}} \rightarrow H_J + U (n_f-1)^2 \equiv J \bm{S}_c \cdot \bm{S}_f + U (n_f-1)^2 .
\label{eq:H_Kondo1}
\end{equation}
The last term is effectively a constraint that enforces that the $f$-level is singly occupied. In the Kondo limit it is therefore natural to describe the remaining low-energy spin degree of freedom of the $f$-electrons in terms of a localized spin-$\frac{1}{2}$ operator $\bm{S}_f$. Doing this on every site of the lattice we are left with the spin-spin interaction term in Eq.~\eqref{eq:H_KLM}.

Alternatively this limit can be studied in terms of the electron and hole operators of the last subsection.\cite{Ostlund2007} This representation has the advantage of having nice transformation properties of charge and spin. A disadvantage is that states with more than one quasi-particle present (outside the triplet sector) at a site will have weight in the high-energy sector. This weight can nevertheless be kept small in the Kondo lattice by being close to the atomic limit $t \ll J$.\cite{Ostlund2007}

We will instead use a less symmetric representation of charge and spin with the basis defined in \eqref{eq:statesenumeration_K}. The advantage of this is that it will allow us to entirely get rid of the high-energy sector by taking one of the operators that span the local Hilbert space (i.e. $c_4^\dagger$) to encode the high-energy excitations. This operator is then straightforwardly eliminated from the periodic Anderson model to generate a low-energy theory that is the Kondo lattice model.

There is considerable freedom in how to assign the 8 states in the low energy sector to the low-energy states in the electron-hole basis. In this work we will only consider transformations such that the operators $c_a^\dagger$ ($a=1 \ldots 3$) are odd with respect to fermion parity. This has the advantage that these operators will be fermionic also in between sites in the Kondo lattice. This implies that the low-energy theory is represented by a theory of interacting fermions on a lattice that can be studied with standard many-body methods. The main disadvantage of this representation is that it is not possible to do this without mixing the spin and charge content of the theory. It is also possible to take one or two of the $c_a^\dagger$'s to be hard-core bosons, but we will not consider these types of transformations in this work.

With this constraint on the allowed transformations we have that: 1) The vacuum with energy $E_s = -3J/4$ is the ground state singlet $|1\ra_K =  |0\ra_s$. 2) The charged states with one electron or one hole and energy $E_c = 0$ maps to the states with one quasi-particle ($|2\ra_K$, $|3\ra_K$, and $|4\ra_K$) or three quasi-particles ($|5\ra_K$). 3) The three triplets with energy $E_t = J/4$ maps to the three states with two quasi-particles ($|6\ra_K$, $|7\ra_K$, and $|8\ra_K$). The Hamiltonian \eqref{eq:H_Kondo1} in this basis is therefore always of the form
\begin{equation}
\label{eq:H_J1}
H_J = 
\frac{J}{4}\bigl[ 1 - (n_1 + n_2 + n_3 -2)^2 \bigr],
\end{equation}
independently of how the states are actually assigned to the three fermions. This Hamiltonian is clearly invariant under a U(3) rotation in the vector space spanned by the three operators $(c_1^\dagger, c_2^\dagger, c_3^\dagger)$, but we will restrict ourselves to the real subgroup O(3) in this paper.

\subsection{At half-filling}
\label{sec:Thalffilling}

At half filling it is desirable to use a representation that is symmetric between electrons and holes, and also as rotationally symmetric with respect to spin as possible. We have found that the following procedure gives quite symmetric representations: 1) We would like the state with three quasi-particles to be a state with zero charge and spin on average. A possible choice is $|5\ra_K = s_5 (e_\da^\dagger - h_\ua^\dagger ) |1\ra_{eh} / \sqrt{2}$, with $s_5 = \pm 1$. This couples the charge and spin content of the excitations, but this is unavoidable in this approach. Note that we have arbitrarily picked a direction of the spin of the electron part of the state. 2) The other three odd fermion parity states can be assigned by demanding that the operators $c^\dagger_1$ and $c^\dagger_3$ are related by time-reversal symmetry, we choose the convention $\mathcal{T} c^\dagger_1 \mathcal{T}^{-1} = -c^\dagger_3$ and $\mathcal{T} c^\dagger_3 \mathcal{T}^{-1} = c^\dagger_1$. A possible assignment of the states is therefore
$|2\ra_K = -(e_\ua^\dagger - h_\da^\dagger ) |1\ra_{eh} / \sqrt{2}$,
$|3\ra_K = (e_\ua^\dagger + h_\da^\dagger) |1\ra_{eh} / \sqrt{2}$, and
$|4\ra_K = (e_\da^\dagger + h_\ua^\dagger ) |1\ra_{eh} / \sqrt{2}$.
3) Because $c^\dagger_1$ and $c^\dagger_3$ are related by time-reversal symmetry $\mathcal{T} |7\ra_K = |7\ra_K$. This implies that $|7\ra_K = -s_7 (|6\ra_{eh}+|8\ra_{eh})/\sqrt{2}$, with $s_7 = \pm 1$, since this is the only time-reversal invariant triplet (see the discussion in Sec.~\ref{sec:time_reversal}). 4) The assignment of the remaining two triplets can be parametrized by an angle $\phi_t$ and a sign $s_6 = \pm 1$ according to 
$s_6 |6\ra_K = \cos(\phi_t) (|6\ra_{eh}-|8\ra_{eh})/\sqrt{2} - \sin(\phi_t) |7\ra_{eh}$ and 
$|8\ra_K = \sin(\phi_t) (|6\ra_{eh}-|8\ra_{eh})/\sqrt{2} + \cos(\phi_t) |7\ra_{eh}$.

The generic transformation is therefore parametrized by the numbers $(s_5 \, s_6 \, s_7,\phi_t)$. Working out the transformation and expanding the original $c$-electron operators in terms of the new fermions we find that they are typically of fifth order. This expansion is most easily performed with the aid of a computer.\cite{Ostlund2007} Only the combinations $(+++,0)$, $(-+-,0)$, and $(--+,\phi_t)$ terminate at third order.
The $(+++,0)$  transformation generates the most symmetric representation:
\begin{eqnarray}
c_{c,\ua}^\dagger &=& \frac{c_1^{\,} + c_1^{\dagger}}{2} + \frac{c_2^{\,} - c_2^{\dagger}}{2} ,
\nonumber \\
c_{c,\da}^\dagger &=& -\frac{c_3^{\,} + c_3^{\dagger}}{2} 
+ \frac{(c_1^{\,} - c_1^{\dagger})(c_2^{\,} + c_2^{\dagger})(c_3^{\,} - c_3^{\dagger})}{2} .
\label{eq:representation1}
\end{eqnarray}
This representation is most easily formulated in terms of Majorana fermions. An alternative elementary derivation of this representation is provided in Sec.~\ref{sec:O3}. The other transformations are equivalent up to a rotation, as an example we give the expression for the representation generated by $(--+,\pi)$:
\begin{eqnarray}
c_{c,\ua}^\dagger &=& \frac{(c_1^{\,} + c_1^{\dagger})(1-2 n_2)}{2} 
+ \frac{(c_2^{\,} - c_2^{\dagger})(1-2 n_1)}{2},
\nonumber \\
c_{c,\da}^\dagger &=& -\frac{c_3^{\,} + c_3^{\dagger}}{2} 
+ \frac{(c_1^{\,} + c_1^{\dagger})(c_2^{\,} - c_2^{\dagger})(c_3^{\,} - c_3^{\dagger})}{2} .
\label{eq:representation2}
\end{eqnarray}

\subsection{Away from half-filling}

Away from half-filling there is no reason to try to enforce a symmetry between electron and hole excitations. Considering the case of hole doping the hole excitations will have lower energy than the electron ones. It is therefore natural to choose two of the low-energy creation operators ($c_1^\dagger$ and $c_3^\dagger$ say) to create the two hole states. With this choice the chemical potential term becomes ($\mu <0$ for hole doping)
\begin{equation}
H_\mu = - \mu (n_c -1) = -\mu (n_2  - n_1 - n_3 + 2 n_1 n_3) .
\end{equation}
To be concrete we define $|2\ra_K = h_\da^\dagger |0\ra_s$ and $|4\ra_K = h_\ua^\dagger |0\ra_s$. It is also convenient to let the creation operators $c_1^\dagger$ and $c_3^\dagger$ to be related to each other by time-reversal symmetry as in the case above. This means that the states $|6\ra_K$, $|7\ra_K$, and $|8\ra_K$ can be parametrized exactly as in Sec.~\ref{sec:Thalffilling}. The electron states are defined via 
$|3\ra_K = \bigl[ \cos(\phi_e) e_\ua^\dagger +  \sin(\phi_e) e_\da^\dagger\bigr] |0\ra_s$ and
$s_5 |5\ra_K = \bigl[ -\sin(\phi_e) e_\ua^\dagger +  \cos(\phi_e) e_\da^\dagger\bigr] |0\ra_s$.
Expanding the $c$-electron operators in terms of the new fermions only two classes of transformations terminate at third order. In both cases $\phi_e = \phi_t$ and the sign structure is $(+++)$ or $(-+-)$. The representations are ($\phi = \phi_t$ and $s=s_7$)
\begin{eqnarray}
c_{c,\ua}^\dagger &=& 
\frac{c_1^{\,}(1-n_3) + s c_1^{\dagger} n_ 3}{\sqrt{2}} 
\nonumber \\
&-& \frac{c_2^{\dagger}}{\sqrt{2}} 
\bigl[  \cos(\phi)  + s \sin(\phi) (c_1^{\,} - s c_1^{\dagger}) (c_3^{\,} - s c_3^{\dagger}) \bigr] ,
\nonumber \\
c_{c,\da}^\dagger &=&
- \frac{c_3^{\,}(1-n_1) + s c_3^{\dagger} n_ 1}{\sqrt{2}} 
\nonumber \\
&-& \frac{c_2^{\dagger}}{\sqrt{2}} 
\bigl[  \sin(\phi)  - s \cos(\phi) (c_1^{\,} - s c_1^{\dagger}) (c_3^{\,} - s c_3^{\dagger}) \bigr] .
\label{eq:representation3}
\end{eqnarray}
We leave the investigation of the representations in Eqs.~\eqref{eq:representation2} and \eqref{eq:representation3} for a later study and will in the following focus on the most symmetric representation of Eq.~\eqref{eq:representation1}.

\section{The {\rm O(3)}-symmetric representation at half-filling}
\label{sec:O3}

In this section we will consider the representation in \eqref{eq:representation1} from another point of view. As discussed in the introduction, it is well-known that it is possible to represent a spin-$\frac{1}{2}$ operator in terms of three species of  Majorana fermions $\mu_a$, $a=1 \ldots 3$. These Majorana fermions are real $\mu_a^\dagger = \mu_a^{\, }$, and independent $\{\mu_a , \mu_b\} = \delta_{ab}$. It is then straightforward to check that the operators
\begin{equation}
S_f^1 = -i \mu_2 \mu_3, \; \;
S_f^2 = -i \mu_3 \mu_1, \; \;
S_f^3 = -i \mu_1 \mu_2,
\label{eq:SfrepMajorana}
\end{equation}
satisfy the angular momentum algebra $[S_f^i , S_f^j] = i \e^{ijk} S_f^k$ for a spin-$\frac{1}{2}$ since $\bm{S}_f^2 = 3/4$. Let us now define
\begin{equation}
\gamma_0 \equiv 2i \mu_1 \mu_2 \mu_3.
\label{eq:gamma0def}
\end{equation}
This is a proper Majorana fermion (with $\gamma_0^2 = 1/2$) that commutes with the spin operator of the $f$-electrons: $[\gamma_0 , \bm{S}_f] = 0$. Therefore we can represent another set of Dirac fermion operators (the $c$-electrons) in terms of $\gamma_0$ and three other Majorana fermions $\gamma_a$, $a = 1 \ldots 3$, as
\begin{equation}
c_\ua^{\, } = \frac{\gamma_1 - i \gamma_2}{\sqrt{2}}, \; \; \; \;
c_\da^{\, } = \frac{-\gamma_3 - i \gamma_0}{\sqrt{2}}.
\label{eq:cdef}
\end{equation}
This representation can be found in e.g. Ref.~\onlinecite{Miranda1994}, but the novelty here is to use the composite operator in \eqref{eq:gamma0def} for $\gamma_0$, instead of an independent Majorana fermion. Both choices satisfy the correct operator algebra, but if one keeps $\gamma_0$ as an independent fermion one will somehow have to deal with the fact that the Hilbert space has been enlarged, see e.g. the discussions in Refs.~\onlinecite{Shastry_Sen_1997} and \onlinecite{Biswas2011}. The model with an independent $\gamma_0$ was studied in the context of odd-frequency pairing at the mean field level in Ref.~\onlinecite{Miranda1994}. The effect of the enlarged Hilbert space then shows up as an additional term in the mean field Hamiltonian, i.e. $H_0$ in their Eq.~(3.3). The remaining part of their mean field Hamiltonian is similar to ours, the main difference being that they do not generate nonlocal hopping terms involving the $\mu$'s.

The $c$-electron spin operators are
\begin{eqnarray}
S_c^1 &=& -i ( \gamma_2 \gamma_3 + \gamma_1 \gamma_0) /2, \nonumber \\
S_c^2 &=& -i ( \gamma_3 \gamma_1 + \gamma_2 \gamma_0) /2, \nonumber \\
S_c^3 &=& -i ( \gamma_1 \gamma_2 + \gamma_3 \gamma_0) /2.
\label{eq:ScrepMajorana}
\end{eqnarray}
In terms of the parity operators $p_a = 2 i \gamma_a \mu_a$ ($a = 1 \ldots 3$), which each has eigenvalues $\pm 1$, the exchange term $H_J = J \bm{S}_c \cdot \bm{S}_f$ can be worked out to be
\begin{multline}
\label{eq:H_J2}
H_J = \frac{J}{8}(p_1 + p_2 + p_3) - \frac{J}{8} (p_1 p_2 + p_2 p_3 + p_3 p_1)
\\
= \frac{J}{8} (1- p_1 p_2 p_3) (p_1 + p_2 + p_3).
\end{multline}
With the identification $p_a = 2 n_a -1$ we see that \eqref{eq:H_J1} and \eqref{eq:H_J2} are equivalent. Note that the spin-spin exchange term has become partly quadratic in the fermions in this representation. This is not the case if one treats $\gamma_0$ as an independent Majorana fermion.

\subsection{Pseudospin symmetry at half-filling}

At half-filling the bipartite Hubbard model, and hence the bipartite symmetric Anderson model possess another symmetry. This is called pseudospin symmetry and is implemented by exchanging the roles of electrons and holes in one of the spin components.\cite{Yang_Zhang_1990,Sigrist_RMP_1997} We can implement this by taking $\gamma_0 \rightarrow - \gamma_0$. The generators of the pseudospin algebra are then
\begin{eqnarray}
I_c^1 &=& -i ( \gamma_2 \gamma_3 - \gamma_1 \gamma_0) /2, \nonumber \\
I_c^2 &=& -i ( \gamma_3 \gamma_1 - \gamma_2 \gamma_0) /2, \nonumber \\
I_c^3 &=& -i ( \gamma_1 \gamma_2 - \gamma_3 \gamma_0) /2.
\end{eqnarray}
It is straightforward to check that $[I^i , I^j] = i \e^{ijk} I^k$ and $[I^i , S^j] = 0$.
If we enforce unit occupancy for the $f$-electrons we see that $\bm{I}_f = 0$ so that the pseudospin algebra of the $f$-electrons becomes trivial. Combining the spin and pseudospin symmetry the system has a global SO(4) symmetry,\cite{Yang_Zhang_1990} which is very transparent in the Majorana representation.\cite{Sen_Shastry_1997}

Going away from half-filling only the generator $I_c^3$ commutes with the Hamiltonian and we recognize $I_c^3$ as the generator of the U(1) gauge symmetry related to charge conservation. This can be seen by including a chemical potential term for the $c$-electrons in the Hamiltonian:
\begin{equation}
H_\mu = - \mu (n_c -1) = i ( \gamma_1 \gamma_2 - \gamma_3 \gamma_0)\mu 
= - 2 \mu I_c^3 .
\end{equation}

\subsection{Arbitrary Kondo lattice}

Let us now consider another site with the same representation:
\begin{equation}
\tilde{c}_\ua = \frac{\tilde{\gamma}_1 - i \tilde{\gamma}_2}{\sqrt{2}}, \; \; \; \;
\tilde{c}_\da = \frac{-\tilde{\gamma}_3 - i \tilde{\gamma}_0}{\sqrt{2}},
\end{equation}
The hopping term between neighboring sites then becomes
\begin{multline}
H_{n.n.} = -t \sum_{\sigma=\da,\ua} (c^\dagger_{\sigma} \tilde{c}_{\sigma} + \text{h.c.}) 
\\
= - i t \Bigl(\gamma^{\,}_3 \tilde{\gamma}_0 + \tilde{\gamma}_3 \gamma_0 +  \tilde{\gamma}_2 \gamma_1 +  \gamma^{\,}_2 \tilde{\gamma}_1  \Bigr).
\end{multline}
This representation has the advantage that it generates at most quartic fermion terms in the Hamiltonian. This implies that the interaction terms can be decoupled using standard Hubbard-Stratonovich transformations.

\subsection{Bipartite Kondo lattice at half-filling -- an O(3)-symmetric representation}
\label{sec:symmericrep}

On a bipartite lattice it is useful to employ different representations on the two sublattices. We introduce an extra phase of $\pi/2$ on one sublattice such that 
\begin{equation}
\tilde{c}_\ua = i \Bigl( \frac{\tilde{\gamma}_1 - i \tilde{\gamma}_2}{\sqrt{2}}\Bigr), \; \; \; \;
\tilde{c}_\da = i \Bigl( \frac{-\tilde{\gamma}_3 - i \tilde{\gamma}_0}{\sqrt{2}}\Bigr).
\end{equation}
Because this just involves a gauge transformation it leads to the same on-site exchange term. Using this the hopping term between two neighboring sites becomes
\begin{equation}
H_{n.n.} =
 -t \sum_{\sigma=\da,\ua} (c^\dagger_{\sigma} \tilde{c}_{\sigma} + \text{h.c.}) 
= -t \sum_{a=0}^3 i \gamma^{\,}_a \tilde{\gamma}_a .
\end{equation}
In the representation in terms of $\gamma_a$ and the $\mu_a$ ($a = 1 \ldots 3$) $i \gamma_0 \gamma_0'$ is a non-linear operator. This implies that we have made a non-linear transformation that preserves a fraction of $3/4$ of the linearity of the hopping term. Explicitly the remaining non-linear part of the hopping term can be written as
\begin{equation}
i \gamma_0 \tilde{\gamma}_0 =
-4 (i \mu^{\,}_1 \tilde{\mu}_1) (i \mu^{\,}_2 \tilde{\mu}_2) (i \mu^{\,}_3 \tilde{\mu}_3).
\label{eq:nonlinearhopping}
\end{equation}
Clearly the Hamiltonian has a global O(3) symmetry that is obtained by the rotation of all local vectors of $(\gamma_1,\gamma_2,\gamma_3)$ and $(\mu_1,\mu_2,\mu_3)$ in the same way. This corresponds to a rotation of spin and pseudospin with the same angle, i.e., the three symmetry generators are $S_c^j + S_f^j + I_c^j$ for $j = 1 \ldots 3$. Note also the peculiar feature that the $c$-electron charge and spin operators are no longer quadratic in fermions in this representation.

\section{Mean field study}
\label{sec:meanfield}

In this section we will use a mean field Hamiltonian, which is quadratic in the fermions, to approximate the interacting fermion theory. For concreteness we will only solve the mean field equations for the 1D lattice with nearest neighbor hopping, but the generalization to other bipartite lattices is straightforward. The Kondo lattice Hamiltonian in the representation of Sec.~\ref{sec:symmericrep} is
\begin{eqnarray}
H_{\text{KLM}} = &-& t \sum_{a=0}^3 \sum_{\la i, j \ra} i \gamma_a(\bm{r}_i) \tilde{\gamma}_a (\bm{r}_j)
\nonumber \\
&+& \sum_i H_J (\bm{r}_i) + \sum_j \tilde{H}_J (\bm{r}_j)
\nonumber \\
&-& \mu 
\sum_{i} \bigl[ i \gamma_2(\bm{r}_i) \gamma_1 (\bm{r}_i) + 
i \gamma_3(\bm{r}_i) \gamma_0 (\bm{r}_i) \bigr]
\nonumber \\
&-& \mu 
\sum_{j} \bigl[ i \tilde{\gamma}_2(\bm{r}_j) \tilde{\gamma}_1 (\bm{r}_j) + 
i \tilde{\gamma}_3(\bm{r}_j) \tilde{\gamma}_0 (\bm{r}_j) \bigr] .
\label{eq:Htbsame}
\end{eqnarray}
$H_J (\bm{r}_i)$ and $\tilde{H}_J (\bm{r}_j)$ are the generalizations of Eq.~\eqref{eq:H_J2} to include a lattice index. The two sublattices are distinguished by the absence or the presence of a tilde.

\subsection{O(3)-symmetric mean field at half-filling}
\label{sec:O3meanfield}

The simplest mean field Hamiltonian at half-filling (i.e. $\mu = 0$) is manifestly O(3) invariant:
\begin{eqnarray}
& & H_{\text{O(3)}} = \sum_{a=1}^3 H_{a}, 
\label{eq:HO3}\\
H_a &=& -t \sum_{\la i, j \ra} i \gamma_a(\bm{r}_i) \tilde{\gamma}_a (\bm{r}_j)
+  \sum_{\la i, j \ra} g(\bm{r}_j - \bm{r}_i) i \mu_a(\bm{r}_i) \tilde{\mu}_a (\bm{r}_j)
\nonumber \\
 &+& V \sum_{i}  i  \gamma_a (\bm{r}_i) \mu_a (\bm{r}_i) 
 + \tilde{V} \sum_{j}  i \tilde{\gamma}_a (\bm{r}_j) \tilde{\mu}_a (\bm{r}_j).
 \label{eq:Hmf_a}
\end{eqnarray}
This form can be motivated from a mean field decoupling of \eqref{eq:Htbsame} using \eqref{eq:H_J2} and \eqref{eq:nonlinearhopping}. We now take the ground state of the mean field Hamiltonian in \eqref{eq:HO3} as a trial state to approximate the ground state of the full interacting theory of Eq.~\eqref{eq:Htbsame}. Rather than fixing the variational parameters by the usual Hartree-Fock decoupling procedure directly, we will keep them arbitrary for the time being, since it is in general possible that different mean field Hamiltonians give the same trial state.\cite{GiulianiVignaleBook}

It is straightforward to diagonalize this problem by going to Fourier space using
\begin{equation}
\gamma_a(\bm{r}_i) = \frac{1}{\sqrt{N/2}} {\sum_{\bm{k}}}' \bigl[
e^{i \bm{k} \cdot \bm{r}_i} \gamma_a^{\,}(\bm{k}) + e^{-i \bm{k} \cdot \bm{r}_i} \gamma_a^\dagger (\bm{k})
\bigr],
\label{eq:fouriergamma}
\end{equation}
and similarly for the other operators.
The prime indicates that one should only include one of the states for each pair of $\bm{k}$ and $-\bm{k}$ in the sum, see e.g. the discussion in Refs.~\onlinecite{Miranda1994,Bulla1997}. This is a consequence of $\gamma^{\,}_a(-\bm{k}) = \gamma_a^\dagger(\bm{k})$. The physics is independent of the choice of $\bm{k}$ or $-\bm{k}$. Note also that the Brillouin zone corresponds to a lattice with two sites per unit cell, hence the $N/2$, where $N$ denotes the total number of lattice sites. Introducing the spinors $\Psi_a(\bm{k}) = [ \mu_a(\bm{k}) , \gamma_a(\bm{k}) , \tilde{\mu}_a(\bm{k}) , \tilde{\gamma}_a(\bm{k})]^{T}$ one has
\begin{equation}
H_{\text{O(3)}}  
= \sum_{a=1}^3 {\sum_{\bm{k}}}' \Psi_a^\dagger(\bm{k}) \Hca(\bm{k}) \Psi_a^{\,}(\bm{k}) ,
\end{equation}
where the Hamiltonian matrix $\Hca(\bm{k})$ is
\begin{equation}
\Hca(\bm{k}) =
\begin{pmatrix}
0 & -i V & i g(\bm{k}) & 0 \\
i V & 0 & 0 & -i t \alpha(\bm{k}) \\
-i g^* (\bm{k}) & 0 &0 & -i \tilde{V} \\
0 & i t \alpha^* (\bm{k}) &i \tilde{V} & 0 
\end{pmatrix} .
\label{eq:Hmatris}
\end{equation}
Here $\alpha(\bm{k}) = \sum_{j} e^{i \bm{k} \cdot \bm{\delta}_j}$ and $g(\bm{k}) = \sum_{j} g(\bm{\delta}_j) e^{i \bm{k} \cdot \bm{\delta}_j}$, where $\bm{\delta}_j$ are the vectors that connects one lattice site (without a tilde) to its nearest neighbors. The mean field Hamiltonian is straightforwardly diagonalized in any dimension, but for simplicity we only perform the mean field analysis in 1D in the following. Setting the nearest neighbor distance to 1 sums can then be converted to integrals with the replacement $\frac{1}{N}{\sum}' \rightarrow \int_0^{\pi/2} \frac{dk}{2 \pi}$. We now introduce $g_\pm = [g(1) \pm g(-1)]/2$, and the mean field solution involves solving for $g_\pm$, $V$, and $\tilde{V}$. For each value of $k$ the spectrum of $\Hca(k)$ is
\begin{equation}
\label{eq:mfnongenericspectrum}
E_{s,t} = \frac{ \pm_s \sqrt{A+B}+ \pm_t \sqrt{A-B} }{2} ,
\end{equation}
where $\pm_s$ and $\pm_t$ are two independent signs and
\begin{eqnarray}
A &=& V^2 +\tilde{V}^2+t^2 \alpha^2 + g_+^2 \alpha^2 + g_-^2 \beta^2 , \nonumber \\
B^2 /4 &=& (V \tilde{V} + t g_+ \alpha^2)^2 + t^2 g_-^2 \alpha^2 \beta^2 , \nonumber \\
\alpha &=& 2 \cos(k) , \; \; \; \; \beta = 2 \sin(k).  
\label{eq:mfspectrum1}
\end{eqnarray}
Taking the ground state of $H_{\text{O(3)}}$ as a trial state the variational ground state energy per site is (taking $\tilde{V} = V$ for simplicity)
\begin{multline}
\e_{var,\text{O(3)}} = - \frac{3 t}{2} \sum_{\text{bonds}}  \la  i \gamma \tilde{\gamma} \ra
+ 2 t \sum_{\text{bonds}}  \la  i \mu \tilde{\mu} \ra^3 
\\
+ \frac{3 J}{4} \la i \gamma \mu \ra - \frac{3J}{2} \la i \gamma \mu \ra^2 .
\label{eq:EvariationalO3}
\end{multline}
Here and throughout the rest of this section we will drop the indexes on the operators since only averages of bilinears with two equal indexes are non-zero because of the O(3) symmetry. Minimizing this variational energy is typically equivalent to a mean field calculation. We have found two classes of solutions to the mean field equations that give low values of the variational ground state energies: one that is translationally invariant and one that is dimerized.

\subsubsection{Translationally invariant phase}

This phase has $\tilde{V} = V > 0$, $g_- = 0$, and  $g_+ \geq 0$. In this case the ground state energy per lattice site of $H_a$ in each component is
\begin{multline}
\e_{0,a} = - \frac{1}{N}{\sum_{\bm{k}}}' \sqrt{4 V^2 + (t + g_+)^2 |\alpha(\bm{k})|^2}
\\ =
- \frac{V E ( - a_+^2 )}{\pi}.
\label{eq:O3varenergy}
\end{multline}
where $a_+ = ( t + g_+ )/V$ and $E(x)$ is the complete elliptic integral of the second kind.\cite{AS_Book} From this we can compute the translational invariant averages of the operators that appear in the variational calculation by differentiation with the result
\begin{eqnarray}
\la i \gamma \mu \ra &=& \la i \tilde{\gamma} \tilde{\mu} \ra = \frac{\partial \e_{0,a}}{\partial V}
= - \frac{1}{\pi} K (-a_+^2) , \nonumber \\
\la i \mu \tilde{\mu} \ra  &=&  - \la i \gamma \tilde{\gamma} \ra = \frac{\partial \e_{0,a}}{\partial t}
= \frac{K(-a_+^2) - E(-a_+^2)}{\pi a_+}.
\label{eq:averagesO3}
\end{eqnarray}
Here $K(x)$ denotes the complete elliptic integral of the first kind.\cite{AS_Book} Interestingly, if we view the ground state of the mean field Hamiltonian as a trial state, the bound on the ground state energy does only depend on the variational parameters $g_+$ and $V$ in the combination $a_+$. This means that there is a one-parameter family of mean field Hamiltonians that have the same ground state. We can imagine to try to fix the best value of $a_+$ in four ways: 1) We use the mean field Hamiltonian to construct a trial density matrix at finite temperature $T$. Taking the limit $T \rightarrow 0$ in the trial free energy the entropy term is maximized if one minimizes the gap in the mean field Hamiltonian. In our system this means that we should choose $g_+ = 0$. 2) We can consider a calculation to second order in the interaction Hamiltonian $H_I = H_{\text{KLM}} - H_{\text{O(3)}}$. 3) We maximize the energy gap in $H_{\text{O(3)}}$ by taking $g_+=t$. This has the additional appealing property that every eigenvalue in \eqref{eq:mfnongenericspectrum} becomes double degenerate since $A=B$ with this choice. 4) We use the standard Hartree-Fock decoupling procedure.\cite{GiulianiVignaleBook}

The usual Hartree-Fock scheme gives
\begin{equation}
V = \frac{J}{4} \bigl( 1 - 4 \la i \gamma \mu \ra \bigr) ,  \; \; \; \;
g_+ = 4 t \la i \mu \tilde{\mu} \ra^2 ,
\label{eq:meanfield_eqs}
\end{equation}
and picks out particular values of $g_+$ and $V$. Note that the expectation value in the atomic ground state gives $V = 3 J/4$ and $g_+ = 0$. The mean field self-consistency conditions \eqref{eq:averagesO3} and \eqref{eq:meanfield_eqs} are easily solved numerically. The variational energy is exact in the limit $J/t \rightarrow \infty$ but gives $-t (3/\pi + 4/\pi^3)$ instead of the correct value $-t 4 /\pi$ if the limit $J/t \rightarrow 0$. In the limit $J=0$ the energy of this state is therefore about  $15 \%$ too high and hence not a good approximation to the ground state. The trial state is better at intermediate values of $t/J$: taking $J=1$ and  $t/J = 1/2$ the best variational energy is $\e_{var,\text{O(3)}} \approx -0.878$, which should be compared to the most accurate estimate from high-order series expansions $\e \approx - 0.926$.\cite{Shi1995} The result is therefore about 5 \% to high for these parameters. The discrepancy can presumably be made smaller by considering fluctuations around the mean field state.

\subsubsection{Dimerized phase}

For smaller values of $J/t$ a dimerized solution, which has the form of a spin-Peierls state, is found to be energetically favorable. It is characterized by $\tilde{V} = V > 0$ and $g_- = \pm (g_+ - \delta)$ with $0 < \delta < g_+ $. The state with $\delta = 0$ is maximally dimerized and has all of its $f$-spins locked up into singlets with one of its nearest neighboring $f$-spins. As $\delta$ grows the dimerization diminishes until it goes away when $g_- = 0$.

For $J/t \gg1$ the optimal value of $g_-$ is found to be extremely small and no gain in energy is found compared with the translationally invariant phase. For $J/t \ll 1$ we find that $\delta \ll g_+$, but a non-zero $\delta$ is needed for a self-consistent solution. Taking the state with $\delta = 0$ as a variational wave function we find that this solution is energetically favorable to the translational invariant phase for $J/t \lesssim 1.24$ in 1D. In the limit $J/t \rightarrow 0$ it gives $-t (3/\pi +1/4)$ which is about $5 \%$ too high, but clearly favorable to the translationally invariant state.

\subsection{SO(2) $\times$ Z$_2$-symmetric mean field at half-filling}

In the O(3)-symmetric mean field states there are three degenerate fermion bands. This leads to a natural description of triplet excitations in terms of two quasi-particle excitations, but it is unnatural in terms of the original description in terms of $c$-electrons and localized $f$-spins. The O(3) symmetry of the Hamiltonian is also broken in the presence of a chemical potential or crystal fields. In this subsection we will therefore allow for a less symmetric mean field solution that has a global SO(2)$\times$Z$_2$ symmetry with the SO(2) generator $S_c^3 + S_f^3+I_c^3$. We can then write the mean field Hamiltonian as $H_{\text{SO(2)}}  = H_{12}+H_{3}$, where $H_3$ is of the same form as in Eq.~\eqref{eq:Hmf_a}.
The mean field Hamiltonian in the remaining components can be decomposed into an onsite part and a hopping part according to $H_{12} = H_{12}^l + H_{12}^t$. The allowed terms in this Hamiltonian are restricted by symmetry. The general on-site local term is
\begin{eqnarray}
H^l_{12} &=& V (i \gamma_1 \mu_1 + i \gamma_2 \mu_2) 
+ \tilde{V} (i \tilde{\gamma}_1 \tilde{\mu}_1 + i \tilde{\gamma}_2 \tilde{\mu}_2) 
 \nonumber \\
&+& m_0 (i \mu_1 \mu_2  + i \gamma_1 \gamma_2 ) 
+ \tilde{m}_0 (i \tilde{\mu}_1\tilde{\mu}_2  + i \tilde{\gamma}_1 \tilde{\gamma}_2 )  \nonumber \\
&+&  m_1 (i \gamma_1 \mu_2 + i \mu_1 \gamma_2 )
+ \tilde{m}_1 (i \tilde{\gamma}_1 \tilde{\mu}_2 + i \tilde{\mu}_1 \tilde{\gamma}_2 ) \nonumber \\
&+& m_3 (i \mu_1 \mu_2 - i \gamma_1 \gamma_2 )  
+ \tilde{m}_3 (i \tilde{\mu}_1 \tilde{\mu}_2 - i \tilde{\gamma}_1 \tilde{\gamma}_2 ) ,
\end{eqnarray}
and the general hopping term is
\begin{multline}
H^t_{12} = -t \sum_{a=1,2} \sum_{\la i, j \ra} i \gamma_a(\bm{r}_i) \tilde{\gamma}_a (\bm{r}_j)
\\
+  \sum_{a=1,2} \sum_{\la i, j \ra} g(\bm{r}_j - \bm{r}_i) i \mu_a(\bm{r}_i) \tilde{\mu}_a (\bm{r}_j)
\\
+  \sum_{\la i, j \ra} h(\bm{r}_j - \bm{r}_i) 
\bigl[ i \mu_1(\bm{r}_i) \tilde{\mu}_2 (\bm{r}_j) - i \mu_2(\bm{r}_i) \tilde{\mu}_1 (\bm{r}_j) \bigr] .
\end{multline}
Note that we use the same notation for some of the variational parameters in $H_3$ and $H_{12}$, i.e. $g$ and $V$, although their values will in general be different.
All in all there are 16 mean field parameters (in 1D), with 4 in the third component, in the most general SO(2)-symmetric mean field Hamiltonian. As in the previous section the spectrum of $H_{12}$ can be found by going to Fourier space using \eqref{eq:fouriergamma} and diagonalizing the resulting 8$\times8$ matrix. Generically the spectrum is then given by the solutions to a quartic equation for the pair $(\alpha,\beta)$ and one for $(-\alpha,\beta)$, but if the quantity
\begin{multline}
f_1 = 
 (m_0 - \tilde{m}_0) (m^2_1 - \tilde{m}^2_1 + m^2_3 - \tilde{m}^2_3 + V^2 - \tilde{V}^2) \\
 -
(m_3 + \tilde{m}_3) (t^2 \alpha^2 - G_+^2 - G_-^2 ) 
,
\end{multline}
vanishes the spectrum is again given by \eqref{eq:mfnongenericspectrum} with different $A$ and $B$. We have also introduced
$G_+ = g_+ \alpha + h_- \beta$, and $G_- = g_- \beta - h_+ \alpha$. We defer a full analysis of the general mean field to a later study. In the following we confine ourselves to the simplified situation with $\tilde{m}_a = -m_a$ for $a=0,1,3$ (to allow for antiferromagnetism) and $\tilde{V} = V$ (leading to sublattice-symmetric on-site $\la i \gamma \mu \ra$) so that $f_1 \equiv 0$. Then the spectrum is given by \eqref{eq:mfnongenericspectrum} with parameters
\begin{eqnarray}
A &=& t^2 \alpha^2 + G_+^2 + G_-^2 + 2 (V^2 + m^2_0 + m^2_1 + m^2_3 )
, \nonumber \\
B^2 /4 &=&  
(V^2 + m_1^2 + m_3^2 -m_0^2 )^2
+ (m_0 + m_3)^2 t^2 \alpha^2
\nonumber \\ &+&
(G_+^2 + G_-^2 ) [ (m_0 - m_3)^2 + t^2 \alpha^2] 
\nonumber \\
&+& 2 G_+ (V^2 - m_1^2) t \alpha + 4 G_- m_1 V t \alpha,
\label{eq:mfspectrum2}
\end{eqnarray}
and the same with $\alpha \rightarrow - \alpha$. This reduces to \eqref{eq:mfspectrum1} when $m_0 = m_1 = m_3 = h_\pm = 0$ and $\tilde{V} = V$. We also note that this spectrum is the same as the mean field spectrum of Ref.~\onlinecite{ZhangYu2000} when $m_1 = h_{\pm} = g_\pm = 0$. The imposed symmetries leads to the following relations for the operator averages
\begin{eqnarray}
\la i \gamma_1 \gamma_2 \ra &=& - \la i \tilde{\gamma}_1 \tilde{\gamma}_2 \ra , \nonumber \\
\la i \mu_1 \mu_2 \ra &=& - \la i \tilde{\mu}_1\tilde{\mu}_2 \ra , \nonumber \\
\la i \gamma_1 \mu_1 \ra &=& \la i \gamma_2 \mu_2 \ra =
\la i \tilde{\gamma}_1 \tilde{\mu}_1 \ra = \la i \tilde{\gamma}_2 \tilde{\mu}_2 \ra , \nonumber \\
\la i \gamma_1 \mu_2 \ra &=& \la i \mu_1 \gamma_2 \ra = 
- \la i \tilde{\gamma}_1 \tilde{\mu}_2 \ra = - \la i \tilde{\mu}_1 \tilde{\gamma}_2 \ra , \nonumber \\
\la i \mu_1 \tilde{\mu}_2 \ra &=& - \la i \mu_2 \tilde{\mu}_1 \ra.
\label{eq:SO2symmetries}
\end{eqnarray}
The relations between the averages on different sublattices can be derived by considering inversion symmetry about a bond in the crystal together with translational invariance.
The variational energy per site is
\begin{multline}
\e_{var,\text{SO(2)}} = - \frac{t}{2} \sum_{\text{bonds}}  
\bigl( 2 \la  i \gamma_1 \tilde{\gamma}_1 \ra + \la  i \gamma_3 \tilde{\gamma}_3 \ra \bigr)
\\
+ 2 t \sum_{\text{bonds}}  \la  i \mu_3 \tilde{\mu}_3 \ra
\bigl( \la i \mu_1 \tilde{\mu}_1 \ra^2 + \la i \mu_1 \tilde{\mu}_2 \ra^2 +  \la i \mu_1 \mu_2 \ra^2
\bigr) \\
+
\frac{J}{2} \la i \gamma_1 \mu_1 \ra +  \frac{J}{4} \la i \gamma_3 \mu_3 \ra
-J \la i \gamma_1 \mu_1 \ra \la i \gamma_3 \mu_3 \ra
\\
- \frac{J}{2} \la i \gamma_1 \mu_1 \ra^2
- \frac{J}{2} \la i \gamma_1 \mu_2 \ra^2
+ \frac{J}{2} \la i \gamma_1 \gamma_2 \ra \la i \mu_1 \mu_2 \ra ,
\label{eq:EvariationalSO2}
\end{multline}
where we have used the symmetries in \eqref{eq:SO2symmetries}. Comparing this with the O(3) case in Eq.~\eqref{eq:EvariationalO3} the difference is the possibility of having nonzero averages for terms involving a coupling between the first and second component on the second and fourth line. Of particular importance is the possibility of having  $\la i \mu_1 \mu_2 \ra \neq 0$ since this will allow the system to take full advantage of the kinetic term in the limit $J \rightarrow 0$.
The operator averages can be calculated by taking the appropriate derivatives of the mean field ground state energy, just like in the O(3) case. The expressions we need are given in the Appendix in Eqs.~\eqref{eq:3average} and \eqref{eq:12average}. We now have all the pieces in place [i.e., Eqs.~\eqref{eq:mfnongenericspectrum}, \eqref{eq:mfspectrum2}, \eqref{eq:EvariationalSO2}, \eqref{eq:3average}, and \eqref{eq:12average}] to perform the variational mean field study.

\subsubsection{Result of the SO(2)-symmetric mean field study}

\begin{figure}[htb]
\centering
\includegraphics[scale=.8]{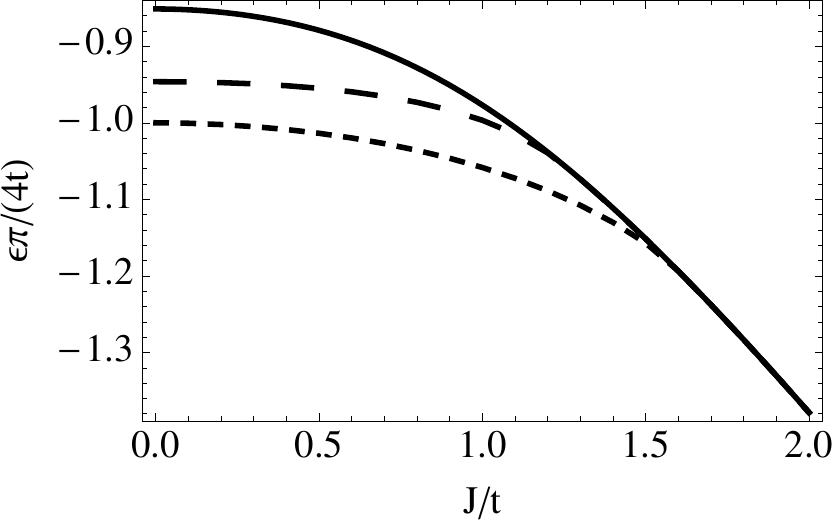}
\caption{Variational energies for different trial states in 1D. From top to bottom: O(3)-symmetric translationally invariant state, O(3)-symmetric dimerized state, and SO(2)-symmetric antiferromagnetic state. Only the last state gives the correct value in the limit $J/t \rightarrow 0$. The mean field theory predicts a transition between the O(3)-symmetric spin liquid phase and an SO(2)-symmetric state with antiferromagnetic order at $J/t \approx 1.56$.}
\label{fig:variationalE}
\end{figure}

The values of $m_1$, both $g_-$'s, and $h_{\pm}$ are found to be extremely small when minimizing the variational energy for all values of $J/t$. Thus we will set these parameters to zero in the following discussion. For large values of $J/t$ the variational parameters flow towards the family of O(3)-symmetric solutions. For $J/t \lesssim 1.56$ nonzero $m_3$ and $m_0$ are found to lower the energy with respect to the O(3) family. This implies an antiferromagnetic SO(2)-symmetric solution that always is favorable to the dimerized O(3)-symmetric solution at the mean field level. This is not surprising in view of earlier work in 1D,\cite{Tsvelik_book} 2D,\cite{Assaad1999} and 3D.\cite{Lacroix_Cyrot_1979} In the limit $J/t \rightarrow 0$ the antiferromagnetic solution reproduces the correct value of the ground state energy, namely $- 4t /\pi$.  In our mean field analysis the transition to the antiferromagnetic state is discontinuous. The variational mean field energies for the different trial states are shown in Fig.~\ref{fig:variationalE}.

\section{Characterizing the O(3)-symmetric phase}
\label{sec:characterization}

In this section we will characterize the O(3)-symmetric mean field phase further by looking at the spin-spin correlation functions and the triplet pairing amplitudes. First we note that the average $c$-electron spin and charge (measured with respect to half-filling) as well as the average $f$-spin are zero on every site. As we will see, the spin-spin correlation functions are rotationally invariant, and we are thus dealing with a spin liquid state. That the system is a gapped spin liquid for large values of $J/t$ is known,\cite{Sigrist_RMP_1997} but the O(3)-symmetric trial state provides a simple realization of such a state for finite values of $t/J$.

\begin{figure}[htb]
\centering
\includegraphics[scale=.82]{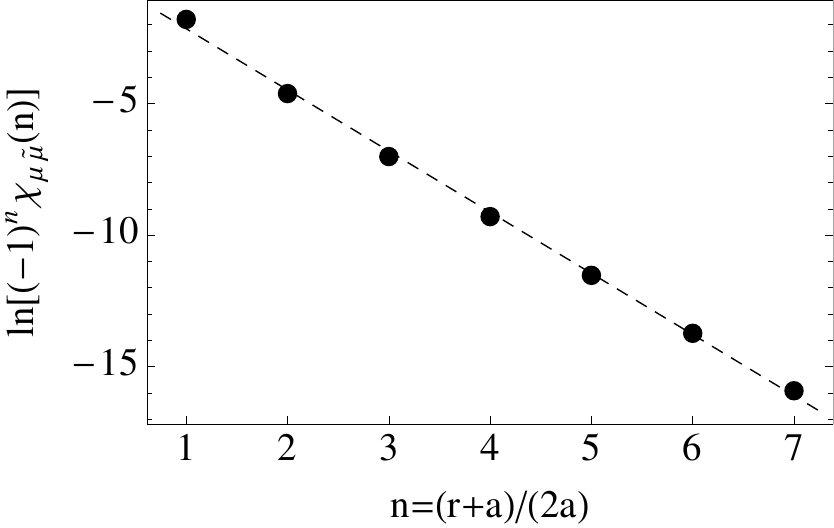}
\caption{The logarithm of the absolute value of the correlation function $\chi_{\mu \tilde{\mu}} (\bm{r}) \equiv \la i \mu_a (\bm{0}) \tilde{\mu}_a (\bm{r}) \ra$ at the value $a_+ = 0.8$, which is appropriate for the trial state at $t/J \approx 0.5$. The dashed line is the fit to a straight line corresponding to exponential decay.}
\label{fig:correlations}
\end{figure}

To characterize the state we first derive the correlation functions for some of the fermion bilinears. Note that we do not consider the trivial autocorrelation functions in the following. Using the Fourier representation in \eqref{eq:fouriergamma} it is straightforward to show that
\begin{multline}
\chi_{\mu \mu} (\bm{r}) \equiv \la i \mu_a (\bm{0}) \mu_a (\bm{r}) \ra 
\\
=  \frac{1}{N/2} {\sum_{\bm{k}}}' \sin (\bm{k} \cdot \bm{r} )
\bigl(1 - 2 \la \mu^\dagger_a(\bm{k}) \mu^{\,}_a(\bm{k}) \ra \bigr) .
\end{multline}
In the translationally invariant O(3) phase $g_- = 0$. In this case particle-hole symmetry enforces $\la \mu^\dagger_a (\bm{k}) \mu^{\,}_a (\bm{k}) \ra = 1/2$ for all $\bm{k}$, which means that $\chi_{\mu \mu} (\bm{r}) = 0$. Similarly $\la i \mu_3 (\bm{r}) \mu_3 (\bm{r}') \ra = 0$ in the SO(2)-symmetric antiferromagnetic state. When the two $\mu$ operators reside on different sublattices we have
\begin{multline}
\chi_{\mu \tilde{\mu}} (\bm{r}) \equiv \la i \mu_a (\bm{0}) \tilde{\mu}_a (\bm{r}) \ra 
\\
=  \frac{1}{N/2} {\sum_{\bm{k}}}' \cos(\bm{k} \cdot \bm{r} )
\la i \mu^\dagger_a(\bm{k}) \tilde{\mu}^{\,}_a(\bm{k}) \ra + \text{h.c.} ,
\label{eq:mutildemucorr}
\end{multline}
in the translationally invariant O(3) phase, and from the diagonalisation of \eqref{eq:Hmatris} we obtain
\begin{equation}
\la i \mu^\dagger_a(\bm{k}) \tilde{\mu}^{\,}_a(\bm{k}) \ra + \text{h.c.}
= - \frac{a_+ \alpha}{\sqrt{4+a^2_+ \alpha^2}} .
\end{equation}
With this result it is straightforward to evaluate the sum in \eqref{eq:mutildemucorr} in the continuum limit numerically. The result is illustrated in Fig.~\ref{fig:correlations} and clearly shows that the result is an alternating almost exponentially decaying function of $r$. In the atomic limit $a_+ \ll 1$, and the non-local correlations are small: $\chi_{\mu \tilde{\mu}} (1) \sim -a_+ / 4$. This will have direct consequences for the spin-spin correlation functions and the triplet pairing amplitudes. Similarly we can show that $\la i \gamma_a (\bm{0}) \gamma_a (\bm{r}) \ra  = 0$, and $\la i \gamma_a (\bm{0}) \tilde{\gamma}_a (\bm{r}) \ra  = - \chi_{\mu \tilde{\mu}} (\bm{r}) $ in the translationally invariant O(3)-symmetric phase.
It is interesting to note that these correlation functions depend on the variational parameters only through $a_+$ and not on the actual spectrum of $H_{\text{O(3)}}$.

\subsection{Spin-spin correlation functions}

In the O(3)-symmetric phase the spin-spin correlation functions are easily shown to be rotationally invariant:
\begin{equation}
\la S_\alpha^i (\bm{r}) S_{\alpha'}^j (\bm{r}') \ra = \delta^{ij} S_{\alpha \alpha'} (\bm{r}' - \bm{r}),
\end{equation}
for all combination of $\alpha, \alpha' = c, f$. This is a consequence of the representations in \eqref{eq:SfrepMajorana}, \eqref{eq:gamma0def}, \eqref{eq:ScrepMajorana}, and the definite fermion parity of all three components in the O(3)-symmetric ground state. It is also possible to work out explicit expressions for the correlation functions in detail, as an example let us look at $f$ spin-spin correlation function. Using the result of the last subsection, $S_{ff} (\bm{r}) = 0$ if the spins reside on the same sublattice and $S_{ff} (\bm{r}) \leq 0$ otherwise. Explicitly we have
\begin{equation}
\la S_f^i (\bm{0}) \tilde{S}_f^j (\bm{r}) \ra = - \delta^{ij} 
\chi^2_{\mu \tilde{\mu}} (\bm{r}).
\end{equation}
From the behavior of $\chi_{\mu \tilde{\mu}} (\bm{r})$ we see that this is a rapidly decaying negative function.

\subsection{Superconducting correlations}

It is easy to see that the singlet Cooper pair amplitude is zero in both the O(3) and the SO(2) phases. This is a consequence of the representation in \eqref{eq:cdef} and \eqref{eq:gamma0def} and the definite fermion parity of the third component in the mean field ground states. The same argument shows that the triplet pairing amplitude $\la c^{\,}_{\ua} (\bm{r}) c^{\,}_{\da} (\bm{r}') + c^{\,}_{\da} (\bm{r}) c^{\,}_{\ua} (\bm{r}') \ra $ vanishes. Using the SO(2) symmetry around the third axis we also find that  $\la c^{\,}_{\ua} (\bm{r}) c^{\,}_{\ua} (\bm{r}') \ra =0$. The third triplet pairing amplitude in the translationally invariant O(3) phase is
\begin{multline}
\Delta_{\da \da} (\bm{r}) \equiv \la c^{\,}_{\da} (\bm{0}) c^{\,}_{\da} (\bm{r}) \ra = 
\\
\la \gamma_3 (\bm{0}) \gamma_3 (\bm{r}) \ra /2
- 2 \la \mu_3 (\bm{0}) \mu_3 (\bm{r}) \ra^3 .
\end{multline}
This expression vanishes in the translationally invariant O(3)-symmetric phase because of the particle-hole symmetry. It remains the last triplet pairing amplitude on different sublattices. In the O(3)-symmetric phase this is given by
\begin{multline}
\Delta_{\da \tilde{\da}} (\bm{r}) \equiv \la c^{\,}_{\da} (\bm{0}) \tilde{c}^{\,}_{\da} (\bm{r}) \ra 
\\
=  \la i \gamma_3 (\bm{0}) \tilde{\gamma}_3 (\bm{r}) \ra /2
+ 2 \la i  \mu_3 (\bm{0}) \tilde{\mu}_3 (\bm{r}) \ra^3 .
\end{multline}
The behavior of $\Delta_{\da \tilde{\da}} (\bm{r})$ is therefore simply related to $\chi_{\mu \tilde{\mu}} (\bm{r})$ and is a rapidly decaying alternating function. The results of this section clearly shows that spin-rotational symmetry, and hence time-reversal symmetry, is broken in a subtle way in the superconducting pairing correlation function in this state.

\section{Conclusions and outlook}
\label{sec:conclusions}

The main finding in this work is an O(3)-symmetric representation of the bipartite Kondo lattice model at half-filling. To the best of our knowledge this representation has not been written down previously. We have used this representation to construct and investigate an O(3)-symmetric mean field state in 1D, and found it to be a good trial wave functions for large to moderate values of $J/t$. At smaller values of $J/t$ a state with antiferromagnetic correlations is favored. The O(3)-symmetric state is a gapped spin liquid with rotationally invariant spin-spin correlations and a finite (short-ranged) triplet pairing amplitude.

For the future it would be interesting to apply the transformation to other lattices than the simplest 1D case considered here, and to see what this representation can tell us about the Kondo lattice away from half-filling. Another direction of research would be to study the effects of the finite triplet pairing amplitude and to allow for slow fluctuations in the direction of this order parameter.

\acknowledgments

We'd like to thank S. \"Ostlund for numerous discussions, C.-Y. Hou for useful comments on the manuscript, and the Swedish research council (Vetenskapsr{\aa}det) for funding.

\appendix

\section{Conventions and mathematical details}

\subsection{Original basis}

We follow the convention of Ref.~\onlinecite{Ostlund2007} to enumerate the states in the original basis:
\begin{eqnarray}
|1\ra &=&  |0\ra, \nonumber \\
|2\ra &=& c_{f,\ua}^\dagger c_{f,\da}^\dagger |0\ra,  \nonumber \\
|3\ra &=& \frac{1}{\sqrt{2}}
(c_{c,\ua}^\dagger c_{f,\da}^\dagger - c_{c,\da}^\dagger c_{f,\ua}^\dagger) |0\ra,  \nonumber \\
|4\ra &=& c_{c,\ua}^\dagger c_{c,\da}^\dagger |0\ra,  \nonumber \\
|5\ra &=& c_{c,\ua}^\dagger c_{c,\da}^\dagger c_{f,\ua}^\dagger c_{f,\da}^\dagger |0\ra, \nonumber \\
|6\ra &=& c_{c,\da}^\dagger c_{f,\da}^\dagger |0\ra, \nonumber \\
|7\ra &=& \frac{1}{\sqrt{2}}
(c_{c,\ua}^\dagger c_{f,\da}^\dagger + c_{c,\da}^\dagger c_{f,\ua}^\dagger) |0\ra,  \nonumber \\
|8\ra &=& c_{c,\ua}^\dagger c_{f,\ua}^\dagger |0\ra, \nonumber \\
|9\ra &=& c_{f,\da}^\dagger |0\ra, \nonumber \\
|10\ra &=& c_{c,\da}^\dagger |0\ra, \nonumber \\
|11\ra &=& c_{c,\da}^\dagger c_{f,\ua}^\dagger c_{f,\da}^\dagger |0\ra, \nonumber \\
|12\ra &=& c_{c,\ua}^\dagger c_{c,\da}^\dagger c_{f,\da}^\dagger |0\ra, \nonumber \\
|13\ra &=& c_{f,\ua}^\dagger |0\ra, \nonumber \\
|14\ra &=& c_{c,\ua}^\dagger |0\ra, \nonumber \\
|15\ra &=& c_{c,\ua}^\dagger c_{f,\ua}^\dagger c_{f,\da}^\dagger |0\ra, \nonumber \\
|16\ra &=& c_{c,\ua}^\dagger c_{c,\da}^\dagger c_{f,\ua}^\dagger |0\ra.
\label{eq:statesenumeration_0}
\end{eqnarray}

\subsection{Time-reversal symmetry}
\label{sec:time_reversal}

We choose the phase convention for the action of the time-reversal operator $\mathcal{T}$ on a generic spin-full fermion level described by the operators $a^{\,}_\da$ and  $a^{\,}_\ua$ to be 
\begin{equation}
\label{eq:timereversal_convention}
\mathcal{T} a^{\dagger}_\ua \mathcal{T}^{-1} = a^{\dagger}_\da ,  \; \; \; \;
\mathcal{T} a^{\dagger}_\da \mathcal{T}^{-1} = -a^{\dagger}_\ua .
\end{equation}
This convention implies that 1) The singlets $|1\ra$ to $|5\ra$ are invariant under time reversal as expected. 2) The $\ua$ states transforms into the corresponding $\da$ states (and vice versa) with our phase convention. 3) The triplets transform as
\begin{equation}
\mathcal{T} | l=1, m \ra = - (-1)^m  | l=1, -m \ra,
\end{equation}
which is different from the transformation of the conventional spherical harmonics.\cite{Sakurai_QM} Restricting ourselves to real coefficients only the triplet $(|6\ra+|8\ra)/\sqrt{2}$ is invariant under time-reversal.

\subsection{Unitary transformations}

Let us first define two different basis sets: $\{ |l\ra_1 \}$ is defined as in Eq.~\eqref{eq:statesenumeration_0} with $c_{c,\sigma}^\dagger \rightarrow c_{c1,\sigma}^\dagger$ and $c_{f,\sigma}^\dagger \rightarrow c_{f1,\sigma}^\dagger$. Similarly for $\{ |l\ra_2 \}$ with $c_{c,\sigma}^\dagger \rightarrow c_{c2,\sigma}^\dagger$ and $c_{f,\sigma}^\dagger \rightarrow c_{f2,\sigma}^\dagger$. The unitary operator $\hat{U}$ implements the transformation between the old basis $\{ | l \ra_1 \}$ and the new basis $\{ | l \ra_2\}$ via the relations 
\begin{equation}
|l\ra_2 = \hat{U} |l\ra_1, \quad \text{for } l=1 \ldots 16.
\end{equation}
In the following we will define $U$ to be the matrix representing $\hat{U}$ in the old basis, i.e.
\begin{equation}
U_{k,l} \equiv {}_1\la k | \hat{U} | l\ra_1 = {}_1\la k | l\ra_2.
\end{equation}
From the definition of the states, and completeness of the basis, we see that the creation operators in the two bases are related by
\begin{equation}
c_{f2,\sigma}^\dagger = \hat{U} c_{f1,\sigma}^\dagger \hat{U}^\dagger , \; \; \; \;
c_{c2,\sigma}^\dagger = \hat{U} c_{c1,\sigma}^\dagger \hat{U}^\dagger .
\end{equation}
This clearly preserves the fermionic anticommuation relations. We also have $c_{f2,\sigma}|0\ra_2 = 0$ iff $c_{f1,\sigma}|0\ra_1 = 0$ etc., which is consistent with the notion that annihilation operators annihilates the vacuum.

\subsection{The electron-hole basis}

The electron-hole basis of Ref.~\onlinecite{Ostlund2007} is obtained by substituting $c_{c,\sigma}^\dagger \rightarrow h_\sigma^\dagger$ and $c_{f,\sigma}^\dagger \rightarrow e_\sigma^\dagger$ in Eq.~\eqref{eq:statesenumeration_0}, and changing the vacuum state to $|0\ra_s$. Explicitly
\begin{eqnarray}
|1\ra_{eh} &=&  |0\ra_s, \nonumber \\
|2\ra_{eh} &=& e_\ua^\dagger e_\da^\dagger |0\ra_s,  \nonumber \\
|3\ra_{eh} &=& \frac{1}{\sqrt{2}}
(h_\ua^\dagger e_\da^\dagger - h_\da^\dagger e_\ua^\dagger) |0\ra_s,  \nonumber \\
|4\ra_{eh} &=& h_\ua^\dagger h_\da^\dagger |0\ra_s,  \nonumber \\
|5\ra_{eh} &=& h_\ua^\dagger h_\da^\dagger e_\ua^\dagger e_\da^\dagger |0\ra_s, \nonumber \\
|6\ra_{eh} &=& h_\da^\dagger e_\da^\dagger |0\ra_s, \nonumber \\
|7\ra_{eh} &=& \frac{1}{\sqrt{2}}
(h_\ua^\dagger e_\da^\dagger + h_\da^\dagger e_\ua^\dagger) |0\ra_s, \nonumber \\
|8\ra_{eh} &=& h_\ua^\dagger e_\ua^\dagger |0\ra_s, \nonumber \\
|9\ra_{eh} &=& e_\da^\dagger |0\ra_s, \nonumber \\
|10\ra_{eh} &=& h_\da^\dagger |0\ra_s, \nonumber \\
|11\ra_{eh} &=& h_\da^\dagger e_\ua^\dagger e_\da^\dagger |0\ra_s, \nonumber \\
|12\ra_{eh} &=& h_\ua^\dagger h_\da^\dagger e_\da^\dagger |0\ra_s, \nonumber \\
|13\ra_{eh} &=& e_\ua^\dagger |0\ra_s, \nonumber \\
|14\ra_{eh} &=& h_\ua^\dagger |0\ra_s, \nonumber \\
|15\ra_{eh} &=& h_\ua^\dagger e_\ua^\dagger e_\da^\dagger |0\ra_s, \nonumber \\
|16\ra_{eh} &=& h_\ua^\dagger h_\da^\dagger e_\ua^\dagger |0\ra_s .
\label{eq:statesenumeration_s}
\end{eqnarray}

\subsection{The electron-hole transformation}

Diagonalizing Eq.~\eqref{eq:H_Anderson1} and using the procedure of Sec.~\ref{sec:transformation_eh} to assign the states we obtain the matrix $U_{eh}$ that implements the transformation from the original basis in \eqref{eq:statesenumeration_0} to the new one in \eqref{eq:statesenumeration_s}. Explicitly
\begin{equation}
\label{eq:Umat_eh}
U_{eh} = 
\begin{pmatrix}
U_s & 0 & 0 & 0 \\
0 & \bm{1}_3 & 0 & 0 \\
0 & 0 & U_\da & 0 \\
0 & 0 & 0 & U_\ua
\end{pmatrix},
\end{equation}
with submatrices
\begin{equation}
U_s = 
\begin{pmatrix}
0 & 0 & 0 & -1 & 0 \\
\sin(\varphi_2)/\sqrt{2}  & 0 & 1/\sqrt{2} & 0 & -\cos(\varphi_2)/\sqrt{2} \\
\cos(\varphi_2) & 0 & 0 & 0 & \sin(\varphi_2) \\
\sin(\varphi_2)/\sqrt{2} & 0 & -1/\sqrt{2} & 0 & -\cos(\varphi_2)/\sqrt{2}  \\
0 & 1 & 0 & 0 & 0 \\
\end{pmatrix},
\end{equation}
\begin{multline}
U_\da = U_\ua \\  =
\begin{pmatrix}
0 & \cos(\varphi_1) & 0 & -\sin(\varphi_1) \\
0 & \sin(\varphi_1) & 0 & \cos(\varphi_1) \\
-\sin(\varphi_1) & 0 & \cos(\varphi_1) & 0 \\
\cos(\varphi_1) & 0 & \sin(\varphi_1) & 0
\end{pmatrix}.
\end{multline}
$\bm{1}_3$ is the $3\times3$ unit matrix, $\tan(\varphi_1) = 2W / (U+\sqrt{U^2 + 4 W^2})$, and
$\tan(\varphi_2) = 4W / (U+\sqrt{U^2 + 16 W^2})$.

\subsection{Basis for the low-energy sector in the Kondo limit}

To describe the Hilbert space in the low-energy sector in the limit $U \rightarrow \infty$ we only need three operators. It is therefore natural to use one operator (i.e. $c_4^\dagger$) to describe excitations in the high-energy sector. We use the following convention to label the states
\begin{eqnarray}
|1\ra_K &=&  |0\ra_s, \nonumber \\
|2\ra_K &=& c_1^\dagger |0\ra_s,  \nonumber \\
|3\ra_K &=& c_2^\dagger |0\ra_s,  \nonumber \\
|4\ra_K &=& c_3^\dagger |0\ra_s,  \nonumber \\
|5\ra_K &=& c_1^\dagger c_2^\dagger c_3^\dagger |0\ra_s, 
\label{eq:statesenumeration_K} \\
|6\ra_K &=& c_2^\dagger c_3^\dagger |0\ra_s,  \nonumber \\
|7\ra_K &=& c_3^\dagger c_1^\dagger |0\ra_s,  \nonumber \\
|8\ra_K &=& c_1^\dagger c_2^\dagger |0\ra_s,  \nonumber \\
| m \ra_K &=& c_4^\dagger (-1)^{n_1+ n_2 + n_3}  |m-8\ra_K,  \; m = 9 \ldots 16. \nonumber
\end{eqnarray}
If we are only interested in the low-energy sector of the theory the assignment of the states in the high-energy sector does not matter. We can therefore make any convenient consistent choice, for example
\begin{eqnarray}
|9\ra_K &=& |5\ra_{eh}, \; \;
|10\ra_K = |11\ra_{eh}, \; \;
|11\ra_K = |12\ra_{eh},
\nonumber \\
|12\ra_K &=& |15\ra_{eh}, \; \;
|13\ra_K = |16\ra_{eh}, \; \;
|14\ra_K = |2\ra_{eh},
\nonumber \\
|15\ra_K &=& |3\ra_{eh}, \; \;
|16\ra_K = |4\ra_{eh} .
\end{eqnarray}
If the high-energy operator $c_4^\dagger$ is of interest one should make a better informed choice.

\subsection{Expressions for the operator averages}

The expressions that we need to calculate the operator averages in the SO(2)-symmetric mean field states are
\begin{eqnarray}
\sum_{\text{bonds}} \la i \gamma_3 \tilde{\gamma}_3 \ra &=&  - 2 \partial_{t} \e_3 ,
\nonumber \\
\la i \gamma_3 \mu_3 \ra &=& \partial_V \e_3 ,
\nonumber \\
\la i \mu_3 \tilde{\mu}_3 (\pm 1) \ra &=& ( \partial_{g_+} \pm \partial_{g_-} ) \e_3 ,
\label{eq:3average}
\end{eqnarray}
for averages in the third component and
\begin{eqnarray}
\sum_{\text{bonds}} \la i \gamma_1 \tilde{\gamma}_1 \ra &=&  - \partial_{t} \e_{12} ,
\nonumber \\
2 \la i \mu_1 \tilde{\mu}_1 (\pm 1) \ra &=& ( \partial_{g_+} \pm \partial_{g_-} ) \e_{12} ,
\nonumber \\
2 \la i \mu_1 \tilde{\mu}_2 (\pm 1) \ra &=& ( \partial_{h_+} \pm \partial_{h_-} ) \e_{12} ,
\nonumber \\
2 \la i \gamma_1 \mu_1 \ra &=& \partial_V  \e_{12} ,
\nonumber \\
2 \la i \mu_1 \mu_2 \ra &=& ( \partial_{m_0} + \partial_{m_3} ) \e_{12} ,
\nonumber \\
2 \la i \gamma_1 \gamma_2 \ra &=& ( \partial_{m_0} - \partial_{m_3} ) \e_{12} ,
\nonumber \\
2 \la i \gamma_1 \mu_2 \ra &=& \partial_{m_1}  \e_{12} ,
\label{eq:12average}
\end{eqnarray}
for averages in the other two. 

\bibliography{Kondorefs}

\end{document}